\documentclass[11pt]{article}
\usepackage{latexsym}
\usepackage{amsmath,amssymb,color}
\usepackage{graphicx}

\usepackage{bbold}

\DeclareMathAlphabet{\mathpzc}{OT1}{pzc}{m}{it}

\newcommand{\ba}{\begin{eqnarray}}
\newcommand{\ea}{\end{eqnarray}}
\newcommand{\be}{\begin{equation}}
\newcommand{\ee}{\end{equation}}
\newcommand{\nn}{\nonumber}

\newcommand{\fun}{\mathbb{\Sigma} }
\newcommand{\funS}{\mathbb{S} }

\def\3s{{s \choose 3}}
\def\4s{{s \choose 4}}
\def\5s{{s \choose 5}}
\def\6s{{s \choose 6}}
\def\12{\frac{1}{2}}

\def\bec{\begin{center}}
\def\ec{\end{center}}

\def\cD{{\cal D}}

\def\nn{\nonumber}

\def\tr{{\rm tr}}

 \def\det{{\rm det\,}}
\def\be{\begin{equation}}
\def\ee{\end{equation}}
\def\bea{\begin{eqnarray}}
\def\eea{\end{eqnarray}}
\def\ba{\begin{array}}
\def\ea{\end{array}}

\begin{document}

\title{Translation invariant time-dependent solutions to massive gravity II}

\author{J.~Mourad\footnote{mourad@apc.univ-paris7.fr} $\;$and D.A.~Steer\footnote{steer@apc.univ-paris7.fr}\\
{\it AstroParticule \& Cosmologie,}\\
{\it UMR 7164-CNRS, Universit\'e Denis Diderot-Paris 7,}\\
{\it CEA, Observatoire de Paris, F-75205 Paris Cedex 13, France}}
\maketitle

{\bf Abstract}
This paper is a sequel to arXiv:1310.6560  [hep-th] and is also devoted to translation-invariant solutions of ghost-free  massive gravity in its moving frame formulation.
Here we consider a mass term which is linear in the vielbein (corresponding to a $\beta_3$ term in the 4D metric formulation) in addition to the cosmological constant. We determine explicitly the constraints, and from the initial value formulation show that  the time-dependent solutions can have singularities at a finite time.  Although the constraints give, as in the $\beta_1$ case, the correct number of degrees of freedom for a massive spin two field, we show that the lapse function can change sign at a finite time causing a singular time evolution. This is very different to the $\beta_1$ case where time evolution is always well defined. We conclude that the $\beta_3$ mass term can be pathological and should be treated with care.

\tableofcontents

\section{Introduction}
\label{sec:intro} 

This paper  is devoted to general translation invariant time-dependent solutions of ghost-free massive gravity (for reviews of massive gravity, see \cite{Rubakov:2008nh,Hinterbichler:2011tt,claudia}), and is a follow up paper to \cite{mejihad} which will hereafter be referred to as [I].
The general framework of massive gravity and the motivations were presented in detail in [I] where one particular mass term --- which is cubic in the vielbien in $D=4$ dimensions, and corresponds to the $\beta_1$ mass term in the metric formulation (see e.g.~\cite{claudia})  --- was considered.  This case was singled out in \cite{dmz1} as allowing a  simple and covariant way of deducing an extra scalar constraint.  It is also singled out in the Hamiltonian analysis, being the only one allowing the constraints to be determined explicitly \cite{Hassan,Kluson,OthersCounting}.

 The general mass term in $D$ dimensions depends on $D-1$ constants $\beta_i$ ($i=1,\ldots,D-1$) in addition to the cosmological constant term. Here we shall consider the one linear in the vielbein, namely $\beta_{D-1}$. In that case, the covariant analysis of \cite{dmz1} shows that a symmetry condition is imposed on the moving frame veilbein components, but does it not lead to an extra scalar constraint. Within the simplified framework of space-independent solutions, we shall show the origin of the necessary extra scalar constraint, and thus determine explicitly all the constraints and the equations of motion.  We show that the time evolution is well posed provided the lapse function $N$ does not vanish.  In fact, the lapse is obtained from the extra scalar constraint and is fully determined by the other fields and their first derivatives. As opposed to the $\beta_1$ case (namely the mass term which depends on $D-1$ factors of the vielbein) in which the lapse function $N(t)$ is strictly positive,  here we show that the sign of $N(t)$ can change.  This leads to singularities in the time evolution which occur at a finite time.  This is the crucial difference with respect to the $\beta_1$ case where the sign of $N$ remains constant.  In [I] we showed that there is a sector in the $\beta_1$-theory which is stable (for related pathologies of massive gravity theories see also \cite{Defelice,Fasiello,Volkov,Deser}): for $\beta_3$-theory this is no longer the case, and singularities are generic for $D>3$.

The paper is organised as follows. In Section \ref{sec:action} we give a brief summary of the moving frame formulation of massive gravity. Section \ref{sec:3}  is devoted to the general analysis of translation-invariant fields. We use a convenient ADM-like decomposition with a lapse function, a shift vector and a symmetric $(D-1)\times (D-1)$ matrix. We show that the Bianchi identities\footnote{The Bianchi identities correspond, in the Hamiltonian language, to the secondary constraints.} together with the constraints  arising from the ${0i}$-components of the equations of motion lead to the vanishing of the shift vectors. This is similar to the $\beta_1$ case and we expect it to be a general property of massive gravity. The Bianchi identities leave one further scalar constraint which once used in the equations of motion --- and in particular after putting these in a form showing their well-posedness --- 
leads to a new scalar constraint.  This extra scalar constraint, which was missing in the analysis of \cite{dmz1}, provides the expression of the lapse function in terms of the symmetric matrix and its first derivative. The equations of motion are well posed provided the lapse function does not change sign.
Contrary to the $\beta_1$ case, this condition is not manifestly true and a case by case study is necessary to prove its validity. This is what we do in Section \ref{sec:4} where we consider some particular cases. In Section \ref{sec:4.1}, we solve analytically the three dimensional case and show that the lapse function is constant. Section \ref{sec:4.2} is devoted to the diagonal solutions in any $D\geq 4$ where all the eigenvalues are equal except one. We show that in that case, for initial conditions in a certain region, $N$ can vanish. This seems to be the most important difference with respect to the $\beta_1$ case and in that respect the latter mass term does not have this pathology. In  the Appendix, we show how using the translation-invariant solutions of this paper and [I] it is possible, by performing a Lorentz transformation, to obtain ``plane wave''  solutions, which can also be seen as the generalisation of the $pp$-waves of general relativity.

\section{Action, equations of motion, and constraints in non-linear massive gravity}
\label{sec:action}

We consider non-linear massive gravity in $D$-dimensions in the vierbein formulation, see \cite{Hinterbichler:2011tt,claudia,dmz1,Hinterbichler:2012cn}. This is described by a dynamical metric $g_{\mu\nu}$ as well as a non-dynamical one $f_{\mu\nu}$, with the corresponding families of 1-forms 
given by $\theta^{A}$ and $f^{A}$ where
\bea
&\eta_{AB}\theta^{A}{}_{\mu}\theta^{B}{}_{\nu}=g_{\mu\nu}\ ,
\label{gdef}
\\
&\eta_{AB}f^{A}{}_{\mu}f^{B}{}_{\nu}=f_{\mu\nu}\ .
\eea
Here the Lorentz indices $A,B=0,\ldots D-1$ are raised and lowered with the Minkowski metric $\eta_{AB}$, 
and the dual vectors $e_{A}$ to the 1-forms $\theta^{A}$ satisfy 
\be
\theta^{A}(e_{B})=\theta^{A}{}_{\mu}e_{B}{}^{\mu}=\delta^{A}{}_{B}.
\label{inv}
\ee

As has been discussed extensively in the literature \cite{Chamseddine:2011mu,Volkov:2012wp,dmz2}, if the symmetry property
\be
e^{C}{}^{\mu} f^{B}{}_{\mu} = e^{B}{}^{\mu} f^{C}{}_{\mu}
\label{sym}
\ee
holds,
then the matrix $g^{-1}f$ has a real square-root, and hence the potential in the metric formulation of massive gravity is well defined. As a result, the action in terms of vierbeins reads \cite{dmz1,Hinterbichler:2012cn,Nibbelink:2006sz}
\be
S=\frac{1}{2}\int \Omega^{AB}\wedge\theta^*_{AB}+\sum_{n=0}^{D-1}\beta_n\int f^{A_1}\wedge\dots\wedge f^{A_n}\wedge\theta^*_{A_1\dots A_n}\ ,
\label{action}
\ee
where the $\beta_n$ are arbitrary parameters, and \cite{DuboisViolette:1986ws} 
\be
\theta^*_{A_1\dots A_n}\equiv{1\over (D-n)!}\epsilon_{A_1\dots A_D}
\theta^{A_{n+1}}\wedge\dots\wedge \theta^{A_{D}}\ 
\ee
is a $(D-n)$-form.  The curvature 2-form $\Omega^{AB}$ is defined by
\be
\Omega^{AB}\equiv d\omega^{AB}+\omega^{A}{}_C\wedge\omega^{CB}\, 
\label{defOmega}\,,
\ee
where the spin-connection $\omega^{AB}$ results from the torsion-free condition $\cD\theta^A\equiv d\theta^A+\omega^{A}{}_B\wedge\theta^B=0$ and the antisymmetry in the indices $A$ and $B$.  From the definition (\ref{defOmega}) the curvature 2-form satisfies the Bianchi identity
\be
\cD\Omega^{AB}\equiv d\Omega^{AB}+\omega^{A}{}_{C}\wedge\Omega^{CB}+\omega^{B}{}_{C}\wedge\Omega^{AC}=0\ .
\ee

We now take (\ref{action}) as our starting point (that is, the condition (\ref{sym}) on the vielbeins is {\it not} imposed). As shown in \cite{dmz1}, for some $\beta_n$ this condition is obtained dynamically, though this is not always necessarily the case (for an example where it does not hold see \cite{ced}). In fact, the class of theories of massive gravity described by (\ref{action}) is larger than that of the metric formulation, and it potentially has a larger space of solutions.  
Furthermore the moving frame formulation does not rely on matrix square roots and is technically much easier to deal with, particularly  concerning  the Bianchi identities which will be extensively used in the following. 
In fact the expression of the derivative of a matrix square root in terms of the derivative of  the matrix is complicated and involves time ordering.

Here we focus on $\beta_0 \neq 0$ and $\beta_{D-1} \neq 0$ in which case (\ref{sym}) is imposed dynamically (see below and \cite{dmz1}).  
The action (\ref{action}) breaks both diffeomorphism and local Lorentz invariance if the non-dynamical vielbein $f^A$ is fixed. In the following we choose $f^{A}\equiv dx^{A}$, which is always possible when $f_{\mu \nu}=\eta_{\mu \nu}$.  The isometry group $SO(1,D-1)$ of the background metric $f_{\mu \nu}$ is a global symmetry group of the theory.  Thus $f^A{}_\mu = \delta^A{}_{\mu}$, and we can identify Lorentz and spacetime indices. Now we define the Einstein tensor as the $(D-1)$-form
\be
G_A\equiv -{1\over 2}\Omega^{BC}\wedge\theta^*_{ABC} \equiv G_{A}{}^{B}\theta^*_B\ ,
\label{defnG}
\ee
where $G_{AB} = R_{AB} - \eta_{AB} R/2$ with $R_{AB} = \Omega_{ACB}{}^C$. Thus
\bea
G_{AB} &=&  e_C{}^\mu\partial _\mu\omega^{C}{}_{AB}- e_B{}^\mu\partial _\mu\omega^{C}{}_{AC} -\omega^{C}{}_{AD}\omega^{D}{}_{BC} +\omega^{D}{}_{AB} \omega^{C}{}_{DC}
\nn
\\
&+&
 \frac{\eta_{AB}}{2}\left[
-\omega^{C}{}_{DE}\omega^{DE}{}_{C} +2 e_I{}^\mu\partial _\mu\omega^{CI}{}_{C} +  \omega^{D}{}_{ED} \omega^{FE}{}_{F} \right]
\label{G1}
\eea
where
\be
\omega^A{}_B = \omega^{A}{}_{BC} \theta^C .
\nn
\ee

Then the field equations following from (\ref{action}) read
\be
G_A=t_A\ ,
\label{eom}
\ee
or equivalently $G_{AB}=t_{AB}$, with
\be
t_A\equiv \sum_{n=0}^{D-1} \beta_n f^{A_1}\wedge\dots\wedge f^{A_n}\wedge\theta^*_{AA_1\dots A_n}\equiv
t_{A}{}^{B}\theta^*_B\ 
\label{tAdef}
\ee
so that 
\be
t_A \wedge \theta_B = (-1)^{D-1} \, t_{AB} \, \varepsilon
\ee
with $\varepsilon$ the volume element $\varepsilon \equiv \theta^0 \wedge \theta^1 \wedge \ldots \wedge \theta^{D-1}$.  For $\beta_0 \neq 0$ and $\beta_{D-1}\neq 0$, it follows that
\be
t_{AB} =  \beta_0 \eta_{AB} + \beta_{D-1} (D-1)! \frac{\theta_{BA}}{\det\theta} .
\label{ihp}
\ee
As mentioned before, Lorentz and space-time indices are identified so that $\theta_{AB} = \theta^C{}_\mu \eta_{CA} \delta_B^\mu 
$.

Notice that as a result of diffeomorphism invariance of the Einstein-Hilbert term, the Bianchi identity
\be
\cD G_A=0=\cD t_A\ 
\label{bb}
\ee
 must hold, whilst Lorentz invariance imposes 
\be
G_{[AB]}=0=t_{[AB]} \ .
\ee
Thus from (\ref{ihp}), the
symmetry of $t_{AB} = t_{BA}$ implies the symmetry of the matrix $\theta_{AB}$. 
In order to guarantee that Minkowski space is a vacuum solution, and that the equations of motion reduce to those of Fierz-Pauli \cite{Fierz:1939ix} in the linearized limit, we choose  $\beta_0$ and $\beta_{D-1}$ to be related to the mass $m$ of the spin 2 field by $\beta_0=m^2 = -\beta_{D-1}(D-1)!$. Thus the 
equations of motion (\ref{eom}) become
\be 
{G_{AB}= m^2\left[\eta_{AB} -  \frac{\theta_{AB}}{\det{\theta}} \right]} \, .
\label{G2}
\ee

Finally \cite{dmz1} the Bianchi identity (\ref{bb}) leads to the $D$ constraints
\be
\omega^B{}_{AC}\theta^{C}{}_B=0
\label{conB}
\ee
or equivalently
\be
\partial_{A}\theta^{B}{}_{B}  - \partial_B \theta^{B}{}_{A} = 0.
\label{conB2}
\ee
As discussed in \cite{dmz1} the various traces of the equation of motion do not lead to a further scalar constraint, contrary to the cases in which only $\beta_1$ or $\beta_2$ are non-zero.   In that respect, the situation we consider here of $\beta_{D-1} \neq 0$ is singled out from those considered before.  

In the following, we will use the simplified framework of translation invariant fields to analyse in detail the equations of motion and find the origin of this extra scalar constraint needed to have the correct number of degrees of freedom.

\section{Equations of motion and constraints}
\label{sec:3}

We now consider solutions invariant under spatial translations and split the symmetric moving frame components $e_{AB}$ and $\theta^{AB}$ into the ADM-type form 
\bea
e_{00}&=&-N\, ,\qquad e_{0i}=-Nn_i \, ,\qquad e_{ij}=\pi_{ij}-Nn_in_j \, ,
\eea
and so from (\ref{inv})
\bea
\theta^{00}&=&-\left({1\over N}-\zeta^{ij}n_in_j \right)\, ,\qquad   \theta^{0i}=-\zeta^{ij}n_j \,, \qquad \theta^{ij}=\zeta^{ij}.
\eea
Here $\zeta_{ij}$ is the inverse of $\pi_{ij}$, and 
the $D(D+1)/2$ variables all depend on $t=x^0$. Notice that $\det\theta
= \frac{\det(\zeta)}{N}$.
 
 In the following, we first write down the constraints arising from the Bianchi identities (\ref{conB2}), and then those coming from the equations of motion (\ref{eom}).  These latter constraints will enable us to show that shift vector $n_i$ vanishes (subsection \ref{subsec:ni}).  As a result the remaining equations of motion take a simplified form, and these are discussed in section \ref{subsec:eofm}.   Finally, we show how the equations of motion and Bianchi identities lead to further scalar constraint which determines the lapse function $N$ in terms of $(\zeta,\partial_t {\zeta})$.

\subsection{Bianchi identities}

The time component of the vector constraint (\ref{conB2}) yields a first constraint (denoted by ${\rm C}_1$ for future reference)
\be
{\rm C_1} \equiv \partial_t (\tr\zeta) = 0  
\label{c1}
\ee
or equivalently $\tr\zeta=c$ for a  constant $c$, while the spatial components give
\be
\partial_t (\zeta^{ij}n_j) = 0\,.
\label{ct}
\ee
We now use the constraints coming from the equations of motion (\ref{eom}) to show that $n_i=0$, so that (\ref{ct}) is trivially satisfied.

\subsection{Vanishing shift, $n_i=0$}
\label{subsec:ni}

The most succinct way of finding the $D$ constraints included in the equations of motion is to proceed as follows. 

Let $G_A^{(s)}$ denote the spatial components of the $D-1$ form $G_A$, that is
\be
G_A^{(s)} \equiv \frac{1}{(D-1)!} G_{A i_1 i_2\ldots i_{D-1}} dx^{i_1} \wedge \ldots \wedge dx^{i_{D-1}}
\label{Gs}
\ee
so that 
\be
G_A = G_A^{(s)} + G_A^{(t)}
\nn
\ee
where the remaining time components, $G_A^{(t)}$, contains one $dt$.  Since we consider time-dependent metrics only, $dG_A = dG_A^{(s)}$ and the Bianchi identity (\ref{bb}) reads $dG_A^{(s)} + \omega_A^{\ B} G_B = 0$.  Thus, at most, 
$G_A^{(s)}$ is first order in time derivatives meaning that the $D$ equations
\be
G_A^{(s)} = t_A^{(s)} 
\label{spaceEinstein}
\ee
are constraint equations.  The second order equations of motion are contained in
\be
G_A^{(t)} = t_A^{(t)} \, .
\label{timeEinstein}
\ee

From (\ref{defnG}) it follows that only the components $\Omega^{AB}{}_{ij}$ of the curvature tensor are needed to determine the constraints, and furthermore from (\ref{defOmega}) we have
\be
\Omega^{AB}{}_{ij} = \omega^{A}{}_{C[i}  \omega^{CB}{}_{j]} \, .
\label{ho}
\ee
The relevant components, $ \omega^{A}{}_{Ci}$, of the spin connexion\footnote{Note that the first term of (\ref{defOmega}) will only contribute time-derivatives, and hence will appear in $\Omega^{AB}{}_{0i}$.}
can be determined from
\be 
\omega_{AB\mu}=\omega_{ABC}\theta^C{}_\mu ={1\over 2}(C_{ABC}+C_{ACB}+C_{CBA})\theta^C{}_\mu \, .
\nn
\ee 
The only non-vanishing components of the structure functions
\be 
C_{AB}{}^C=e_{[A}{}^\mu\partial_\mu e_{B]}{}^\nu\theta^C{}_\nu \, = -e_A{}^\mu e_B{}^\nu \partial_{[\mu} \theta_{\nu]}{}^C
\nn
\ee
are
$C_{0i}{}^0 = \partial_t (\zeta n)_i$ and $C_{0i}{}^k = -\partial_t{\zeta}_{i}{}^k$. Thus 
\be
\omega_{0ki} = 
 - \frac{N}{2} \fun_{ki} \, ,
\qquad
\omega_{jki} = 
 \frac{N}{2} (\fun^T )_{i[j} n_{k]}
 \nn
 \ee
where
\be
\fun \equiv \{ \partial_t{\zeta},\pi \} \zeta .
\nn
\ee
Finally, substituting into (\ref{ho}) gives
\bea
\Omega^{qp}{}_{ij} &=& \frac{N^2}{4} \left\{ (1-n^2) \fun^q{}_{[i} \fun^p{}_{j]} + \fun^q{}_{[i} (\fun^T n)_{j]} n^p +  (\fun^T n)_{[i}\fun^p{}_{j]} n^q \right\} \, ,
\label{OmegaS}
\\
\Omega^{0q}{}_{ij} &=& \frac{N^2}{4} \left\{  (\fun^T n)_{[j}\fun^q{}_{i]}\right\} \, .
\label{OmegaT}
\eea

From (\ref{defnG}) as well as (\ref{OmegaS}) and (\ref{OmegaT}), we then obtain
\be
G_p^{(s)} =G_0^{(s)} n_p 
\nn
\ee
with
\be
G_0^{(s)} = \frac{N^2}{8} (\det \zeta) \left[ \tr(\funS^2) -\tr(\funS)^2  - n^T (\funS^2 -\funS\tr(\funS))n \right]  {\rm d}V
\nn
\ee
where ${\rm d}V=dx^1\wedge \ldots\wedge dx^{D-1}$ is the spatial volume element, and the symmetric matrix $\funS$ is defined by
\be
\funS \equiv \{ \partial_t{\zeta},\pi \}
\label{funSdef}
\ee
so that $\fun = \funS \zeta$.
The RHS of (\ref{spaceEinstein}), is determined directly from (\ref{tAdef})  and we find
\bea
t_p^{(s)} &=& m^2 n_p (\det \zeta) {\rm d}V 
\nn
\\
t_0^{(s)} &=& m^2 \left[(\det \zeta)  - 1 \right] {\rm d}V \, .
\nn
\eea
Hence, the we finally arrive at the $D$ constraints contained in $G_A^{(s)} = t_A^{(s)}$: 
\bea
\frac{N^2}{8} \left[ \tr(\funS^2) -\tr(\funS)^2   - n^T (\funS^2 -\funS\tr(\funS))n \right]  n_p &=& m^2 n_p 
 \label{F1}
 \\
\frac{N^2}{8} \left[ \tr(\funS^2) -\tr(\funS)^2   - n^T (\funS^2 -\funS\tr(\funS))n \right]   &=& m^2 \left[1  - \frac{1}{(\det \zeta) } \right] \, .
 \label{F2}
 \eea
On substituting (\ref{F2}) into (\ref{F1}) it follows that
\be
n_i = 0
\label{nizero}
\ee
whilst the remaining scalar constraint reads
\be
\frac{N^2}{8} \left[ \tr(\funS^2) -\tr(\funS)^2 \right] = m^2 \left[1  - \frac{1}{(\det \zeta) } \right]\, .
 \label{finalSSS}
\ee

\subsection{Equations of motion}
\label{subsec:eofm}

The constraint $n_i=0$ which we have found above enormously simplifies the  analysis.  In this subsection we write down the remaining dynamical Einstein equations, namely (\ref{timeEinstein}).  

To do so, it is convenient to change variables.  Notice that
the dynamical metric now takes the form 
\bea
ds^2 &=& g_{\mu \nu}dx^\mu dx^\nu
\nn
\\
& =& - \frac{1}{N^2} dt^2 + (\pi^{-2})_{ij} dx^i dx^j \, ,
\nn
\\
&\equiv& -dT^2 + (B^{-2})_{ij}dx^i dx^j
\label{Tt}
\eea
where $B \equiv \pi^2$ and the new time coordinate $T$ 
is unambiguously determined from (\ref{Tt}) only if $N$ does not vanish at some finite $t=t_\star$.  We will return to this important point and its interpretation below.\footnote{Recall, [I], that for $\beta_1 \neq 0$ the constraints always imply that $N>1$; in the case of $\beta_{D-1}$ we will see that $N$ can vanish for $D>3$.}  In the following we denote by a dot a derivative with respect to $T$
\be
\cdot \equiv \partial_T =  N \partial_t .
\label{Tdef}
\ee
On using
\be
N \funS =- \zeta \dot{B} B^{-1} \pi,
\nn
\ee
(which follows from the definition of $\funS$ in (\ref{funSdef})), the second constraint ${\rm C}_2$ in (\ref{finalSSS}) reads
\be
{\rm C_2} \equiv \left(\tr (\dot{B} B^{-1})\right)^2 - \tr\left( (\dot{B} B^{-1})^2\right) + 8m^2 (1-\det{\pi}) =0 .
\label{c2B}
\ee

The equations of motion for $B(T)$, are given by the $(ij)$-component of the Einstein equation. From (\ref{G1}) and (\ref{G2}) they read
\bea
-\partial_T(\dot{B}B^{-1}) +\partial_T \tr(\dot{B}B^{-1}) \mathbb{1}+ \frac{1}{2}(\dot{B}B^{-1}) \tr(\dot{B}B^{-1})&&
\nn
\\
 - \frac{\mathbb{1}}{4} \left[ \left(\tr (\dot{B} B^{-1})\right)^2 + \tr\left( (\dot{B} B^{-1})^2\right)\right] &=&-2m^2 \left[\zeta(N\det\pi) - \mathbb{1}\right]\, 
,
 \label{eofmleB}
 \eea
and we will refer to them as E (standing for `equations of motion') in the following.
Notice that even when they are written in terms of the new time coordinate $T$, these equations depend explicitly on the lapse function $N(T)$ which is still undetermined. 
As we will discuss in the next subsection, E and ${\rm C}_1$ together in fact generate a third constraint $\rm{C}_3$ which will determine $N(T)$.

 Before doing so, let us discuss the main properties of the equations E.
 First notice that the LHS of (\ref{eofmleB}) can be simplified by using (\ref{c2B}) to eliminate the second term in the square bracket. Then on
defining 
\be
\epsilon \, e^f \equiv \det \pi \, ,  \qquad \epsilon = \frac{\det \pi}{|\det \pi|} = \pm 1,
\label{fdef}
\ee
the equations E become
\be
\frac{1}{2} \partial_T(e^{-f} \dot{B} B^{-1}) = \partial_T(e^{-f} \dot{f})\mathbb{1} +m^2 \left[ (\epsilon - 2 e^{-f}) \mathbb{1}  + N\zeta \epsilon \right] .
\label{useful}
\ee
Second, note that the antisymmetric part of this equation yields the  {\it conserved} matrix 
\be
\gamma \equiv \frac{1}{2} e^{-f} [\dot{B},B^{-1}] ,
\ee
with
$\dot{\gamma}=0$.  Thus the system contains $(D-1)(D-2)/2$ conserved quantities, which correspond to the expected conservation of angular momentum since the system is invariant under spatial rotations.  These will be used extensively below when we construct exact solutions.

Third, on taking the trace of (\ref{useful}) we find
\be
N\beta \epsilon \, (\tr\zeta) = (D-2)\partial_T(e^{-f} \dot{f}) +m^2 (D-1) (\epsilon -  2 e^{-f})
\label{NB}
\ee
which, when substituted back into (\ref{useful}) can be used to eliminate $N$. Indeed we then find a unique traceless 2nd order equation for $B$ which is independent of $N(T)$, namely
\be
 \frac{1}{2} \partial_T(e^{-f} \dot{B} B^{-1}) - \partial_T(e^{-f} \dot{f}) \left[\mathbb{1}  - (D-2)\frac{\zeta}{\tr(\zeta)}\right] = m^2 (\epsilon - 2 e^{-f})  \left[\mathbb{1}  - (D-1)\frac{\zeta}{\tr(\zeta)}\right] .
 \label{useful2}
 \ee
{\it If} this equation is well posed --- something which is not manifest and will be discussed below --- it follows that it can be used to find $\ddot{\zeta}$ in terms of lower derivatives of $\zeta$. As a result, the first term on the right hand side of  (\ref{NB}) can be expressed in terms of $(\dot{\zeta},\zeta)$ thus eliminating all second derivatives from (\ref{NB}) which then becomes a constraint. This will be done in detail in the next subsection.
Notice that in the $\beta_1$ case, see [I], the analogue of (\ref{NB}) is already a constraint since in that case the analogue of ${\rm C}_1$ is $\dot{f}=0$ so that $N$ is determined directly in terms of $\zeta$.

Before writing out this third constraint $\rm{C}_3$ explicitly, it is useful introduce a final set of variables which simplify the equations and constraints further. Motivated from (\ref{useful}) define
\bea
\partial_u \equiv  e^{-f} \partial_T  \, , \qquad C \equiv  e^{-2f} B \, , 
\label{uCdef}
\eea
with $ ' = \partial_u$.
Then the conserved quantity $\gamma$ and the equation of motion E become
\bea
\gamma &=& \frac{1}{2}[C',C^{-1}]\, , 
\label{gammaC}
\\
\partial_u[C^{-1}C']&=&2m^2 e^{-f}[N\zeta \epsilon-(2e^{-f}-\epsilon)\mathbb{1} ]\, ,
\label{eofmCa}
\eea
whilst the constraints ${\rm C}_1$ and ${\rm C}_2$ become
\bea
\tilde{\rm C}_1&\equiv & (\tr\zeta)' = 0 \, ,
\label{first}
\\
\tilde{\rm C}_2&\equiv & \tr\left( ({C'} C^{-1})^2\right) - 4(D-2) f'^2 - 8m^2(e^{-2f}-\epsilon e^{-f})=0.
 \label{second}
\eea
Eqs.~(\ref{NB}) and (\ref{useful2}) in turn read
\bea
-m^2 {(\tr \zeta)}  N\epsilon&=&  (D-2)f'' e^f -m^2 (D-1) (2e^{-f}-\epsilon) \, ,
\label{df}
\\
\partial_u (C^{-1} C')  + 2 \frac{\zeta}{\tr \zeta}(D-2)f'' &=& 2m^2 e^{-f} (\epsilon - 2e^{-f}) \left[ \mathbb{1} - (D-1)\frac{\zeta}{\tr \zeta} \right] \, .
 \label{eofmC}
\eea

Fourth, in these new variables it is straightforward to verify the compatibility of the constraints $\tilde{\rm C}_1$ and $\tilde{\rm C}_2$ with the equation of motion. To do so, we first multiply (\ref{eofmC})  by $C^{-1}C'$ and then take the trace. This gives
\bea 
{1\over 2}\partial_u\left[ \tr(C^{-1}C')^2 \right]&-&
{4(D-2)\over \tr\zeta} f''\left[ (\tr\zeta)'+f'\tr\zeta \right] \qquad \qquad \qquad \qquad\qquad \qquad\qquad 
 \nonumber
\\
&=& 2m^2 e^{-f}(\epsilon-2e^{-f})\left\{ 2f'+{2(D-1)(\tr\zeta)'\over \tr\zeta} \right\} \,
\label{E22}
\eea 
which can be rewritten as
\be
\partial_u( \tilde{\rm C}_2) = - 8m^2 (\tr\zeta)' e^{-f} N \epsilon \, .
\label{geneve}
\ee
Thus we deduce that if $\tilde{\rm C}_2$ is valid for all $u$ and the equations of motion hold, then $\tilde{\rm C}_1$ follows provides $N \neq 0$.   In the Hamiltonian language, this is expressed as $\tilde{\rm C}_1$ being a secondary constraint, with  $\tilde{\rm C}_2$ a primary constant.    
An alternative way of interpreting (\ref{geneve}) is that if $\tilde{\rm C}_1$ holds for all $u$ and  $\tilde{\rm C}_2$ is true at an initial $u$, then must be true for all $u$.

Finally let us verify that Fierz-Pauli theory is recovered in the linearized limit, $\pi = \mathbb{1} + h$ and $B =  \mathbb{1}  + 2h$.  In that case, the first constraint $\tilde{\rm C}_1$ in equation (\ref{first}) gives $\tr{h'}=0$ whilst from $\tilde{\rm C}_2$ in (\ref{second}) it follows that $\tr h = 0$.  Thus $\det \pi=1$ so that from (\ref{fdef}) that $f=0={f'}={f''}$. Also $\tr(\zeta)=(D-1)$, so that from (\ref{df}), $N=1$.  Thus $u=T=t$ and finally the equation of motion (\ref{eofmC}) reduces to
\be
\frac{d^2 h_{ij}}{dt^2} +m^2 h_{ij} = 0
\ee
as required.

\subsection{The third constraint}
\label{sec:dof}

In the previous subsection we obtained $N$ in terms of the second derivative of $f$ (or equivalently $\zeta$), see (\ref{df}).
We also have the equation of motion (\ref{eofmC}). If this equation is well posed it follows that it can be used to find ${\zeta''}$ in terms of lower derivatives of $\zeta$. As a result, the first term on the right hand side of  (\ref{df}) can be expressed in terms of $({\zeta'},\zeta)$ thus eliminating all second derivatives from (\ref{df}) which then becomes a constraint.  

Alternatively, we can determine $N$ in terms of $(\zeta,{\zeta'})$ directly from the observation that the equations of motion (\ref{eofmC}) combined with the first constraint $\tilde{\rm C}_1$ yield a 3rd scalar constraint $\rm{C}_3$.  Schematically the reason is the following: in terms of $\zeta$, Eq.~(\ref{eofmC}) can be rewritten in the form ${\zeta''} = g(\zeta,\dot{\zeta},N)$ for some (matrix) function $g$ which will be determined below but which is clearly linear in $N$.  On taking the trace of this equation of motion, it follows that the left hand side must vanish by $\tilde{\rm C}_1$. Thus we are left with a third constraint $\rm{C}_3$ namely  $\tr(g(\zeta,{\zeta'},N))=0$, which a priori determines $N$ giving $N = N(\zeta,{\zeta'})$.  (As mentioned above, in the $\beta_1 \neq 0$ case [I], 
the analogue of (\ref{eofmC}) contains no terms in first and second derivatives of $\zeta$ and directly determines $N=N(\zeta)$.)

The crucial step is to show the well posedness of the the equations of motion (\ref{eofmC}) --- that is to put them in a form in which it is manifest that ${\zeta}''$ is determined in terms of lower derivatives.  To carry out this procedure, we write the real symmetric matrix $\zeta(u)$ in the form
 \be
\zeta(u)=\sum_{i=1}^{D-1} e^{-\Delta_i(u)}|v_i(u)\rangle\langle v_i(u)|,
\label{evalueszeta}
\ee
where the $|v_i(u)\rangle$ are its orthonormal eigenvectors with eigenvalues\footnote{From now on we only consider positive eigenvalues so that $\epsilon=+1$} $e^{-\Delta_i}$.  Thus from (\ref{fdef}) we have
$f = \sum_i \Delta_i$,
and from (\ref{uCdef}) $C$ is given by $C = \pi^2 e^{-2f}$.  The constraint $\tilde{\rm C}_1$ in (\ref{first}) then implies that
\be
(\tr \zeta)'' = 0 \qquad \Longleftrightarrow \qquad \sum_i \Delta_i'' e^{-\Delta_i} =  \sum_i  \Delta_i'^2 \, e^{-\Delta_i},
\label{C1Delta}
\ee
whilst the conserved anti-symmetric matrix $\gamma$ 
defined in (\ref{gammaC}) is given by
\bea 
\gamma&=& \frac{1}{2} [{C'},C^{-1}] = 
2\sum_{i,j}\sinh^2{(\Delta_i-\Delta_j)}\langle{v_i}(u)|{v'_j}(u)\rangle|v_j(u)\rangle\langle v_i(u)|
\label{FSP}
\eea
so that
\be 
\langle{v_i}(u)|{v'_j}(u)\rangle={\langle v_j(u)|\gamma|v_i(u)\rangle\over 2\sinh^2{(\Delta_i-\Delta_j)}} \,  \qquad (i \neq j).
\label{bet}
\ee
The matrix elements of $\gamma$ in a time independent basis give $(D-1)(D-2)/2$ constants of motion. If we choose that basis to be the eigenvectors $|v_a(0)\rangle$at the initial time $u=0$, these are given by
\bea 
\gamma_{ab}&=&2\sum_{i,j}\sinh^2{(\Delta_i-\Delta_j)}\langle{v_i}(u)|{v'_j}(u)\rangle\langle v_a(0)|v_j(u)\rangle\langle v_i(u)|v_b(0)\rangle\nonumber\\&=&2\sinh^2{(\Delta_a-\Delta_b)(0)}\langle v_b(0)|{v'_a}(0)\rangle. 
\label{gammab}
\eea

We can now proceed to find the third constraint $\rm{C}_3$.  The $(ii)$ components of equations of motion (\ref{eofmCa}) reduce to 
\bea
2 ({\Delta''_i} - f'') -A_i&=&2m^2e^{-f} \left[Ne^{-\Delta_i} +(1-2e^{-f}) \right]
\label{eqm1}
\eea
where
\bea
A_i &\equiv&  \frac{1}{2}\sum_{j \neq i}\sinh{2(\Delta_i-\Delta_j)}\left( 
{\langle v_j(u)|\gamma|v_i(u)\rangle\over \sinh^2{(\Delta_i-\Delta_j)}}
\right)^2 , 
\label{Aidef}
\eea
so that $\sum_i A_i =0$.   The $(ij)$ components of the equations of motion are a consequence of the constraints (as in the $\beta_1$ case [I]).  On summing (\ref{eqm1}) over $i$, we find
\be
f'' = -\frac{m^2 e^{-f}}{(D-2)} \left[ N\tr\zeta + (D-1)(1-2e^{-f})\right]
\nn
\ee
which is of course identical to (\ref{df}).  Then substituting into (\ref{eqm1}) gives
\be
 {\Delta''_i} = \frac{ A_i}{2} + \frac{m^2 e^{-f}}{(D-2)} \left[ N \left((D-2)e^{-\Delta_i} - \tr\zeta \right) - (1-2e^{-f})\right] \, .
 \label{dofm}
 \ee
Finally, the constraint $\rm{C}_3$ comes from combining (\ref{dofm}) with (\ref{C1Delta}): 
\be
{\rm C_3} \equiv {m^2 e^{-f} N} = \frac{\sum_i  \Delta_i'^2 e^{-\Delta_i}+ \frac{m^2 \tr\zeta }{D-2}e^{-f}(1-2e^{-f}) -\frac{1}{2}\sum_i A_i e^{-\Delta_i} 
}{\tr(\zeta^2) - \frac{(\tr\zeta)^2}{D-2}} .
\label{C3}
 \ee
Notice that in the limit $\Delta \rightarrow 0$ then $ \tr\zeta\rightarrow D-1$, and $N\rightarrow 1$ as it should from the linear analysis above.\footnote{In the Hamiltonian language, this constraint $\rm{C}_3$ is expressed as the determination of the Lagrange multiplier $N$.}
For completeness, in terms of the variables $\Delta_i$ and $|v_i\rangle$, the first two constraints read
\bea
\tilde{\rm C}_1 &\equiv& -\sum_i \Delta'_i e^{-\Delta_i}=0
\label{c1c1}
\\
\frac{1}{4}\tilde{\rm C}_2 &\equiv& \sum_i {\Delta'_i}^2 - \Big(\sum_i \Delta'_i\Big)^2 +\frac{1}{4} \sum_{i\neq j} \left( {\langle v_j(u)|\gamma|v_i(u)\rangle\over \sinh^2{(\Delta_i-\Delta_j)}}\right)^2 + 2m^2 e^{-f} (1-e^{-f}) = 0 \, .
\label{c2c2}
\eea

Now we are in a position to analyse the initial value formulation. A combination of (\ref{C3}) and (\ref{dofm}) shows that the problem is well posed provided $N$ does not diverge.
Indeed, suppose that at $u=0$, we are given $\Delta_i(0)$, ${\Delta'_i}(0)$, $|v_i(0)\rangle$ and 
$|{v'_i}(0)\rangle$ that satisfy the two constraints (\ref{c1c1}) and (\ref{c2c2}) with $\gamma$ given by (\ref{gammab}).
From these we obtain $\Delta_i(\delta u)$, $|v_i(\delta u)\rangle$. From (\ref{bet}) we determine $|{v'_i}(\delta u)\rangle$ and from Eqs.~(\ref{dofm}) and (\ref{C3}) we get ${\Delta'_i}(\delta u)$ {\it provided} $N$ does not diverge. Thus, as long as $N$ remains finite, our system of equations are sufficient to solve the system completely once a correct set of initial values satisfying the constraints is given.  It is possible that $N$ diverges at a finite $u$, since the denominator in (\ref{C3}) can vanish consistently with the constraints $\tilde{\rm C}_1$ and $\tilde{\rm C}_2$.

To finish the problem, we of course need to transform back to the original time variable $t$, defined from (\ref{Tdef}) and (\ref{uCdef}) by
\be
\frac{dt}{du} =  e^{-f(u)} N(u)
\label{ttou}
\ee
with $N(u)$ determined from (\ref{C3}).  In order to be able to determine $u(t)$, $N(u)$ should not change sign. Indeed, if $N(u)$ were to change sign then $t$ would not be a monotonic function of $u$ and so we could not invert to find $t(u)$.  From (\ref{C3}) notice that while the first term in the numerator of $N(u)$ is definitely positive, the remaining two terms do not have a definite sign. Hence generically nothing appears to guarantee that $N$ can never vanish.  Below we give an example where, indeed, $N(u)$ changes sign.

\section{Examples}
\label{sec:4}

To understand whether $N$ can change sign, study instabilities etc, it is instructive to search for exact solutions of the equations of motion E and constraints ${\rm C}_1$, ${\rm C}_2$, $\rm{C}_3$.  We will consider two simple situations: $D=3$ dimensions, and $D$-dimensional space-time but vanishing $\gamma$.

Before doing so, notice that due to the constraints, phase space $(\pi_{ij},{\pi}'_{ij})$ contains the $D(D-1)-2$ degrees of freedom necessary to describe a massive spin 2 field, and of those, $(D-1)(D-2)/2 + 1$ are constants of motion.

\subsection{Solutions in $D=3$ dimensions}
\label{sec:4.1}

In $D=3$ the system is integrable.  It is simplest to proceed by writing $\pi(u)=\pi_0(u)+\pi_1(u)\sigma_1+\pi_3(u)\sigma_3$, where $\sigma_i$ are the Pauli matrices, so that 
\be
\det \pi = (\pi_0^2-\pi_1^2-\pi_3^2) \, .
\ee
Then from $\tilde{\rm C}_1$, 
\be
\tr\zeta= c = 2\frac{\pi_0}{\det \pi}
\label{claude}
\ee
where $c$ is a constant, and from the conserved matrix
$\gamma$ in (\ref{gammaC})
\be
\gamma_{12} \equiv L = 
c^2  { [{\pi'_1}\pi_3-{\pi'_3}\pi_1]} \, 
\ee
with $dL/du=0$.
The constraint equation $\tilde{\rm C}_2$ in (\ref{second}) reads 
\be
(\det\pi) ({\pi'_1}^2+{\pi'_3}^2-{\pi'_0}^2) + \left(\frac{L}{c^2}\right)^2 = m^2(1-\det \pi) \, .
\label{eofm}
\ee
One can check that the equation of motion (\ref{eofmC}) is the derivative of this equation.

In order to solve (\ref{eofm}) we change variables to
\be 
\pi_0=\rho\cosh\xi,\qquad \pi_1=\rho\sinh\xi\sin\theta,\qquad\pi_3=\rho\sinh\xi\cos\theta,
\label{snuff}
\ee
so that $\det \pi = \rho^2
= \left(\frac{2}{c} \cosh \xi \right)^2$
and $L=  {\theta'}  \sinh^2 (2\xi)$.  Then (\ref{eofm}) reads
\be
 {\xi'}^2+V_{\rm eff}(\xi) = 0,\label{ene3}
 \ee
 with
\be
V_{\rm eff} = \frac{1}{4} \left[ \frac{L^2}{4\sinh^2\xi} -m^2 c^2 \left( \frac{c^2}{4\cosh^2\xi} - 1 \right) \right] \, .
\label{Vp}
\ee
This is the equation of motion for a particle moving in 1 dimension with an effective potential (\ref{Vp}).
Notice that  as $\xi \rightarrow \pm \infty$, $V \rightarrow m^2 c^2/4$, and hence the field dynamics is bounded in a region of finite $\xi$.
When $L=0$, the potential is negative at $\xi=0$ only for $c\geq 2$.  When $L\neq 0$ there is a potential barrier at $\xi=0$ meaning that $\xi$ cannot change sign during the evolution.  In fact it is straightforward to show that $V_{\rm eff}<0$ only if 
\be
L < m c (c-2) .
\label{req}
\ee
In fact the solution can be obtained exactly, and is given by
\be
\xi(u) = {\rm arcsinh} \left(\left[ \frac{E^2}{2B^2} + \left(\frac{E^4}{4B^2} - F^2 \right)^{1/2}\frac{1}{B}\sin(2B(u-u_0)) \right]\right)^{1/2}
\label{xiu}
\ee
where
\be
F=\frac{L}{4} \, , \qquad B=\frac{mc}{2} \, , \qquad E=\sqrt{ \frac{m^2 c^4}{16} - F^2 - B^2} \, ,
\ee
and $u_0$ determines the initial value of $\xi$.
As expected the motion is periodic in terms of the variable $u$.  

However, we must check whether or not $N(u)$ changes sign during the evolution.  To do so, we use the third constraint $\rm{C}_3$ given in (\ref{C3}).  From (\ref{snuff}), the two eigen-values and vectors of $\zeta$ (see (\ref{evalueszeta})) are given by
\be
e^{-\Delta_1} = \frac{c}{2}\frac{e^{-\xi} }{\cosh\xi} \, , \qquad e^{-\Delta_2} = \frac{c}{2}\frac{e^{\xi} }{\cosh\xi} \, ,
\qquad
|v_1\rangle = \left( \begin{array}{c}
\cos(\theta/2) \\ \sin(\theta/2)
\end{array}\right) \, ,
\qquad
|v_2\rangle = \left( \begin{array}{c}
\sin(\theta/2) \\ - \cos(\theta/2) 
\end{array}\right)
\nn
\ee
so that from its definition in (\ref{Aidef}) it follows that
\be
A_1 = L^2 \frac{\cosh 2\xi}{\sinh^3(2\xi)} = - A_2 \, .
\ee
On using (\ref{Vp}) we finally find from  (\ref{C3}) that $N$ is given by
\be
N = \frac{c}{2} \left( 1 - \frac{L^2}{m^2 c^4} \right).
\label{N3}
\ee 
Thus for $L$'s satisfying the requirement (\ref{req}), $N$ is positive and constant for all $\xi$.
Finally,  in terms of the original time coordinate $t$,
\be
t = \int \frac{N(u)}{\det\pi} du = \frac{Nc^2}{4} \int \frac{1}{\cosh^2(\xi(u))} du
\ee
where $\xi(u)$ is given in (\ref{xiu}).  This can be integrated exactly, but the answer is not particularly illuminating.  Crucially $t$ is well defined and monotonic, and $\xi$ is also periodic in $t$.

As we now show, however, the situation is very different in $D>3$ dimensions.

\subsection{The diagonal case, $\gamma=0$, in $D$ dimensions}
\label{sec:4.2}

We now study the diagonal case
\be
\zeta={\rm diag}(e^{-\Delta_1},\dots, e^{-\Delta_{D-2}}, e^{-\Delta}) 
\label{gammazero}
\ee
where $ \Delta = \Delta_{D-1}$. Thus $\gamma=0$, and the equations of motion and constraints ${\rm C}_1$, ${\rm C}_2$, $\rm{C}_3$ given in (\ref{dofm})-(\ref{c2c2}) can be solved numerically.   It is useful to get some analytic understanding of the dynamics by focusing on the simple case in which all but one of the $\Delta_i$ are identical\footnote{The easiest particular case is the one in which all the $\Delta_i$ for $i=1,\dots D-1$ are equal, so that $\pi$ is proportional to the identity. However, it is straightforward to see that in that case $\pi$ must be the identity itself.}, thus
\be
\Delta_a(u) = \delta(u) \qquad {\rm for} \qquad a = 1\ldots D-2 \, .
\nn
\ee
We have checked that the generic solutions show similar properties to the ones we discuss below.

\begin{figure*}
\includegraphics[width=0.45\textwidth,height=.35\textwidth]{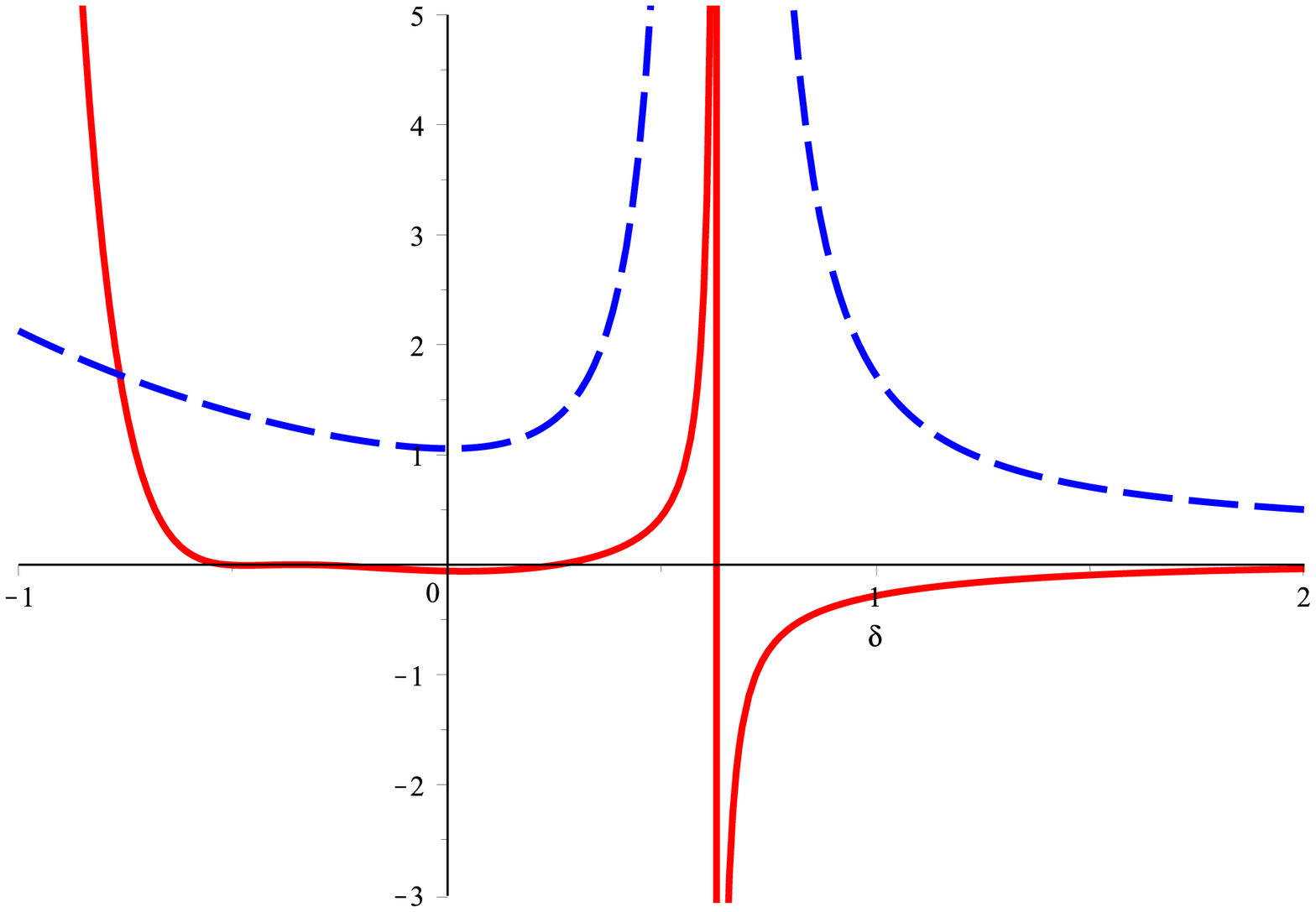}
\hspace{1cm}
\includegraphics[width=0.45\textwidth,height=.35\textwidth]{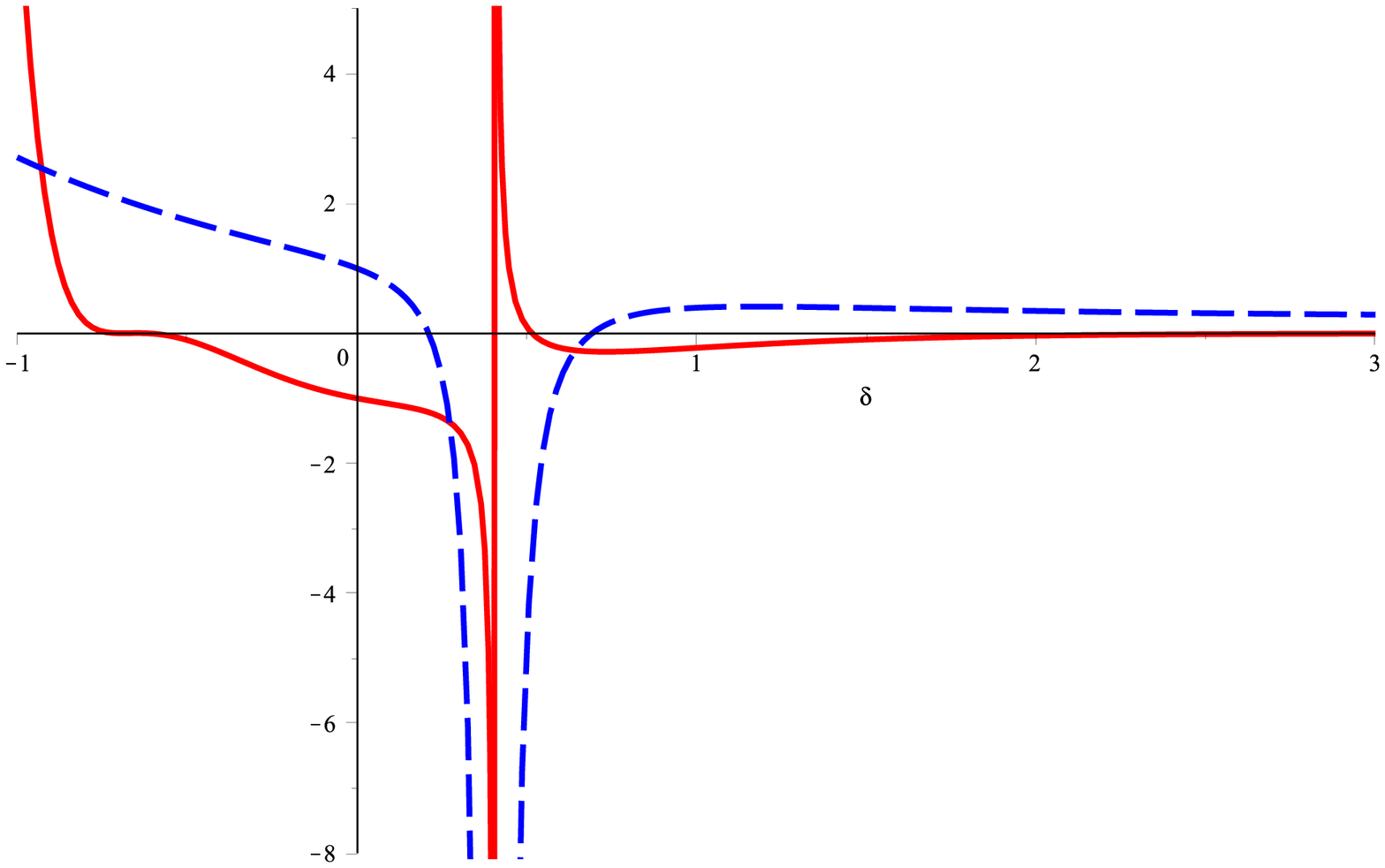}
\caption{The potential $V(\delta)$ (red, solid lines) and $N(\delta)$ (blue dashed lines) in $D=4$ dimensions for which $c_{\rm crit} \simeq 3.78$. LH panel:  $c=3.3<c_{\rm crit}$; RH panel $c=4>c_{\rm crit}$.}
\label{fig:pot}
\end{figure*}

In that case, the equations of motion and constraints directly determine $\delta$ since one can use the constraint $\tilde{\rm C}_1$ to eliminate $\Delta$:
\be 
 \tr \zeta = (D-2)e^{-\delta} + e^{-\Delta}=c,
\label{gogo}
\ee
where $c$ is a constant.   Before doing so, notice that the previous condition implies that $e^{-\delta}$ is bound in the range
\be
0 \leq e^{-\delta} \leq \frac{c}{D-2}\, .
\ee
We then find that
the constraint (\ref{c2c2}) reduces to\footnote{One can check that the equation of motion (\ref{eofmC}) is the derivative of equation (\ref{Ed4}).}
\be
\delta'^2 + V(\delta)=0
\label{Ed4}
\ee
where
\bea
V &=&-2m^2 e^{-(D-2)\delta}\left( \frac{\left(c-(D-2)e^{-\delta}\right)^2}{D-2} \right) \frac{ \left[1 - c \,e^{-(D-2)\delta}+(D-2)e^{-(D-1)\delta} \right]}
{ \left\{ c(D-3)-(D-2)(D-1)e^{-\delta}\right\} }  \, .
\eea
(In $D=3$ dimensions, this potential agrees with the one of the previous section when we set $L=0$ and on correctly identifying $\delta$ and $\xi$.) In order to have a perturbative solution for $\delta \ll 1$ we require $V(0) <0$, which leads to
\be
\frac{(D-1)(D-2)}{D-3} > c >  D-1 \, .
\label{rangec}
\ee
Notice also that
\be
V(\delta \rightarrow \infty) \rightarrow 0 \, ,\qquad V(\delta \rightarrow -\infty)  \rightarrow \infty
\ee
and furthermore that for $D>3$, even though the system has no generalised angular momentum (that is $\gamma=0$ here), the potential diverges at $\delta = \delta_*$ where 
\be
e^{-\delta_*} = \frac{c(D-3)}{(D-2)(D-1)} <  \frac{c}{(D-2)}\,.
\ee
As a result of this potential barrier, the evolution of $\delta$ is divided into disconnected sectors depending on the initial conditions. Furthermore, in each sector we need to determine $N$ which, from (\ref{C3}), is given by
\be
N = \frac{ c \, e^{-(D-2)\delta}  \left[ (D-3) e^{(D-2)\delta} -2c(D-3)+2(D-2)^2 e^{-\delta} \right]}{ (e^{-\delta} - e^{-\delta_*})^2  (D-2)^2 (D-1)^2} \, .
\ee
 (When $D=3$ this reduces to $N=c/2$ as it should from (\ref{N3}) with $L=0$.) Notice that $N$ also diverges at $\delta_*$.  Whether or not $N$ can change sign is determined by the square brackets in the numerator which we denote by $f(e^{-\delta})$.  This function $f$ has a minimum at
 \be
 e^{-\delta_0} = \left(\frac{D-3}{2(D-2)} \right)^{\frac{1}{D-1}} \, 
 \ee
where it takes the value
 \be
 f_{\rm min}  =-2(D-3)[c-c_{\rm crit}],
 \ee
 with
 \be
  c_{\rm crit } = \frac{D-1}{2^{1/(D-1)}} \left( \frac{D-2}{D-3}\right)^{\frac{D-2}{D-1}} \, 
  \ee
  which lies in the range given by (\ref{rangec}).  (For instance, in $D=4$ dimensions $6\leq c\leq 3$ with $c_{\rm crit} = (54)^{1/3} \simeq 3.78$.)
 Furthermore, 
  \be
  f(e^{-\delta_*}) = -2c\frac{D-3}{D-1} \left[ 1-\left(\frac{c_{\rm crit}}{c}\right)^{D-1} \right] \, .
  \ee
  Thus for $c<c_{\rm crit}$ $N$ is always positive, while for $c>c_{\rm crit}$ $N$ is negative in a range of values of $\delta$ around $\delta_0$ which includes $\delta_*$. Figure \ref{fig:pot} shows this generic behaviour as well as that of $V$ for $D=4$.

To conclude, for $c>c_{\rm crit}$, $N$ necessarily vanishes at a finite time $u_*$ so that the system is badly defined. For $c<c_{\rm crit}$ there is well defined evolution if $\delta$ is initially in the minimum of $V$ near $\delta=0$. However, if initially $\delta>\delta_*$ then depending on the sign of $\delta'$, either $\delta \rightarrow \delta_*$ in a finite $u$ (and correspondingly infinite time $t$), or $\delta \rightarrow \infty$ as $u \rightarrow \infty$.

A numerical study of the generic system in $D=4$ in which all the eigenvalues of $\zeta$ are initially different (with $\gamma=0$, see (\ref{gammazero})) shows that the behaviour of the system is qualitatively similar.  Crucially there are initial conditions for which $N$ changes sign in a finite time.

\section{Conclusion}

 In this paper we have studied massive gravity in its vielbein formulation, considering the mass terms parametrised by the parameters $\beta_0$ and $\beta_{D-1}$ only.  In this case, the covariant analysis of \cite{dmz1} showed that a symmetry condition is imposed on the moving frame veilbein components, but does it not lead to an extra scalar constraint.   By focusing on  time-dependent and spatially translational invariant metrics --- which in the context of Fierz-Pauli would correspond to studying plane waves, and in the context of General Relativity to Bianchi I and Kasner solutions ---  we were able to determine the origin of this extra scalar constraint which is required for the theory to have the correct $D(D-1)-2$ degrees of freedom needed to describe a massive spin 2 particle. 
  
We  carried out the analysis using a convenient ADM-like decomposition of the veilbein, with a lapse function $N$, a shift vector $n_i$  and a symmetric $(D-1)\times (D-1)$ matrix $\pi_{ij}$ (or equivalently its inverse $\zeta_{ij}$). In section \ref{subsec:ni} we showed that as a result of the Bianchi identities and the constraints  coming from the ${0i}$-components of the equations of motion, $n_i=0$.  In terms of the variable $u$, which is linked to the original time coordinate $t$ through (\ref{ttou}), we also showed how the Bianchi identities give rise to the scalar constraint (which was denoted by ${\rm C}_1$). This, once used in the equations of motion, was shown to lead to the new scalar constraint, $\rm{C}_3$, which was missing in the analysis of \cite{dmz1}. As discussed in section \ref{sec:dof}, a crucial step in the determination of $\rm{C}_3$ was to put the equations of motion in a form showing their well-posedness, that is with $\zeta''$ explicitly expressed in terms of its lower derivatives.  In passing we also showed that of the $D(D-1)-2$ degrees of freedom, $(D-1)(D-2)/2 + 1$ are constants of motion (coming from the conserved anti-symmetric matrix $\gamma$ as well as $\tr \zeta$).

A notable difference with the $\beta_1$ case studied in [I] is that the equations of motion are well posed provided the lapse function does change sign (nor diverge).  While this was guaranteed in the $\beta_1$ case (where $N(t) \geq 1$ $\forall t$), it is not longer manifestly true for the $\beta_{D-1}$ mass term considered here. A case by case study was needed to prove its validity: in Section \ref{sec:4.1}, we solved analytically the theory in $D=3$ and showed that the lapse function is constant and hence the theory is well defined. However, as shown in section \ref{sec:4.2} this no longer holds in $D=4$ where for initial conditions in a certain region, $N$ can change sign at a finite time $t$, leading to singular time evolution.  In this respect, an important conclusion is that in $D=4$ dimensions the $\beta_3$ mass term can be pathological and should be treated with care.

Finally, in the Appendix, we have shown how the translation-invariant solutions presented here and in [I] can be generalised by performing a Lorentz transformation. As a result we obtain ``plane wave''  solutions, which can also be seen as the generalisation of the $pp$-waves of general relativity.

In the future it would be interesting to use the intuition gained here to understand how to obtain the constraint $\rm{C}_3$ for a general space-time metric, and hence complete the covariant Lagrangian approach of \cite{dmz1} to this $\beta_3$-case.  

 \section*{Acknowledgments}
 We thank C.~Deffayet, J.~Madore, K.~Noui and G.~Zahariade for useful discussions.  DAS thanks the Yukawa Institute for Theoretical Physics at Kyoto University,
where this work progressed significantly during the 
the International Molecule-type Workshop ``Modified gravity'' (YITP-T-13-08).  She is also grateful to CERN for hospitality whilst this work was being finished.

\section{Appendix: Plane waves}

In this appendix we show how, from the time-dependent solutions presented above, we can obtain space- and time-dependent solutions. These are the generalisation of the plane waves of Fierz-Pauli theory, and can also be seen as the generalisation of the $pp$-waves of general relativity.

As we have taken $f^A = dx^A$ from the start, the theory is globally Lorenz invariant and hence it is sufficient to perform a boost on the time-dependent solutions, thus mapping the rest-frame $D$-vector $p_0^\mu = (m,\vec{0})$ to a general momentum $D$-vector $p^\mu = \Lambda^{\mu}_{\; \; \nu} p_0^\nu$ with 
\be 
p^2=-m^2.
\ee
Recall that for a general $D$-vector $v$, we define the longitudinal and  transverse components by
\be 
v_L=-{v.p\over m^2}p, \qquad v_{T}=v-v_L
\ee
with $v_T.p=0$,
and similarly for a general $D$-tensor $t^{\mu\nu}$, its transverse and longitudinal parts are given by
\be 
t_{LL}^{\mu\nu}={p^\mu p^\nu\over m^4}p_\alpha p_{\beta}t^{\alpha\beta},\qquad
t_{LT}^{\mu\nu}=-{p^\mu p_{\alpha}\over m^2}\Delta^{\nu}{}_\beta t^{\alpha\beta},\qquad
t_{TT}^{\mu\nu}=\Delta^\mu{}_\alpha\Delta^\nu{}_\beta t^{\alpha\beta},
\ee
where
 \be 
\Delta_{\mu\nu}=\eta_{\mu\nu}+{p_\mu p_\nu\over m^2}
\ee
and 
$t=t_{TT}+t_{TL}+t_{LT}+t_{LL}$.

Define the  $D$ polarisation vectors $\epsilon^{(A)}(p)_\mu,\ A=0,\dots D-1$  by
\be
\epsilon^{(A)}(\Lambda p_0) = \Lambda \epsilon^{(A)}(p_0)
\ee
where
\be
\epsilon^{(\nu)}(p_0)_\mu =\delta^\nu_{\; \; \mu} \, .
\ee
These form a basis with
\be 
\epsilon^{(A)} \cdot \epsilon^{(B)}=\eta^{AB},\quad \epsilon^{(0)}={p\over m} \, .
\ee
The plane wave solution for the moving frame $\theta^A = \theta^{A}_{\; \; \mu}dx^\mu$ is then  explicitly given in terms of the translation-invariant solutions by
\be 
\theta_{LL}^{\mu\nu}(x)=N(-x\cdot\epsilon^{(0)})\epsilon_{(0)}^{\mu}\epsilon_{(0)}^{\nu}, \qquad
\theta_{TL}^{\mu\nu}=0,\ \qquad \theta_{TT}^{\mu\nu}=\pi_{ij}(-x\cdot\epsilon^{(0)})\epsilon^{(i)\mu}
\epsilon^{(j)\nu}.
\ee
In particular the full solution for the metric $g_{\mu \nu}$ is given by
$g_{\mu\nu}(x) = \eta_{AB}\theta^{A}{}_{\mu}(x)\theta^{B}{}_{\nu}(x)$
where $\theta^{AB} = \theta_{LL}^{AB} +  \theta_{TT}^{AB}$.  Thus, to conclude
\be
g_{\mu\nu}(x) = -N^2(-x\cdot\epsilon^{(0)}) \frac{p_\mu p_\nu}{m^2} + (\pi^2)_{ij} \epsilon^{(i)}_\mu \epsilon^{(j)}_\nu 
\ee
where $N$ and $\pi_{ij}$ can be replaced by any of the time-dependent solutions determined above.

\end{document}